\pdfoutput=1
\documentclass[aip,jcp,amsmath,amssymb,preprint,comment]{revtex4-1}

\usepackage{bm}
\usepackage{dcolumn}
\usepackage{hyperref}
\usepackage{graphicx}
\graphicspath{{figures/}}

\usepackage{amsfonts,amsmath,amssymb}

\usepackage[names]{xcolor}
\usepackage{multirow}

\usepackage{listings}

\newcommand{\be}{\begin{equation}}
\newcommand{\ee}{\end{equation}}
\newcommand{\bea}{\begin{eqnarray}}
\newcommand{\eea}{\end{eqnarray}}

\newcommand{\bvr}{\mathbf{r}}

\begin{document}

\title{Comparison of permutationally invariant 
polynomials, neural networks, and
Gaussian approximation potentials in representing water interactions 
through many-body expansions} 

\author{Thuong T. Nguyen}
\affiliation{Department of Chemistry and Biochemistry, University of California, 
San Diego, La Jolla, California 92093, United States}
\affiliation{San Diego Supercomputer Center, University of California, 
San Diego, La Jolla, California 92093, United States}

\author{Eszter Sz\'ekely}
\affiliation{Engineering Department, University of Cambridge, Trumpington Street, Cambridge CB2 1PZ, United Kingdom}

\author{Giulio Imbalzano}
 \affiliation{Laboratory of Computational Science and Modeling, Institute of Materials, {\'E}cole Polytechnique F{\'e}d{\'e}rale de Lausanne, 1015 Lausanne, Switzerland}%

\author{J\"org Behler}
\affiliation{Universit\"at G\"ottingen, Institut f\"ur Physikalische Chemie, Theoretische Chemie, Tammannstr. 6, 37077 G\"ottingen, Germany}

\author{G\'abor Cs\'anyi}
\affiliation{Engineering Department, University of Cambridge, Trumpington Street, Cambridge CB2 1PZ, United Kingdom}

\author{Michele Ceriotti}
 \affiliation{Laboratory of Computational Science and Modeling, Institute of Materials, {\'E}cole Polytechnique F{\'e}d{\'e}rale de Lausanne, 1015 Lausanne, Switzerland}%

\author{Andreas W. G\"otz}
\affiliation{San Diego Supercomputer Center, University of California, 
San Diego, La Jolla, California 92093, United States}

\author{Francesco Paesani}
\email{fpaesani@ucsd.edu}
\affiliation{Department of Chemistry and Biochemistry, University of California, 
San Diego, La Jolla, California 92093, United States}
\affiliation{San Diego Supercomputer Center, University of California, 
San Diego, La Jolla, California 92093, United States}


\begin{abstract}
The accurate representation of multidimensional potential energy surfaces is a 
necessary requirement for realistic computer simulations of molecular systems.
The continued increase in computer power accompanied by advances in
correlated electronic structure methods nowadays enable routine calculations 
of accurate interaction energies for small systems, which can then be used
as references for the development of analytical potential energy functions (PEFs) 
rigorously derived from many-body expansions.
Building on the accuracy of the MB-pol many-body PEF, we investigate here
the performance of permutationally invariant polynomials, neural networks, 
and Gaussian approximation potentials in representing water two-body and three-body interaction energies, denoting the resulting potentials PIP-MB-pol, BPNN-MB-pol, and GAP-MB-pol, respectively. 
Our analysis shows that all three analytical representations exhibit similar levels of accuracy 
in reproducing both two-body and three-body reference data as well as interaction energies of small water clusters
obtained from calculations carried out at the coupled cluster level of theory,
the current gold standard for chemical accuracy.
These results demonstrate the synergy between interatomic potentials formulated in terms of a many-body expansion, such as MB-pol, that are physically sound and transferable, and machine-learning techniques that provide a flexible framework to approximate the short-range interaction energy terms.
\end{abstract}

\maketitle

\section{Introduction}\label{sect:intro}

Since the first Monte Carlo (MC)\cite{metropolis_n_1953, wood_ww_1957} 
and molecular dynamics (MD)\cite{alder_bj_1959, alder_bj_1960}
simulations of molecular systems, 
computer simulations have become a powerful tool for molecular sciences, complementing 
experimental measurements and often providing insights that are difficult to obtain by other means. 
Although the first simulations were performed for idealized molecular systems,
it was recognized since the beginning that both realism and predictive power of a computer simulation
are directly correlated with the accuracy with which the underlying molecular interactions are described.  

In this context, computer modeling of water is perhaps the most classic example. 
Given its role as life's matrix,\cite{ball_p_2008}  it is not surprising that numerous molecular models 
of water have been developed 
(see Refs.~\citenum{guillot_b_2002, vega_c_2011, shvab_i_2016, cisneros_ga_2016} for recent reviews)
since the first simulations performed by Barker and Watts,\cite{barker_ja_1969} 
and Rahman and Stillinger.\cite{rahman_a_1971} 
However, despite almost 50 years have passed since these pioneering studies, the development of
a molecular model that correctly reproduces the behavior of water from the gas to the condensed
phase still represents a formidable challenge.

From a theoretical standpoint, the energy of a system containing $N$ water molecules can be formally 
expressed through the many-body expansion of the interaction energy (MBE) as\cite{mayer_je_1940}
\begin{equation} \label{eq:mbexpansion}
	E_{\text{N}}(1,\dots,N) = \sum_{i=1}^N V^{\text{1B}}(i) + \sum_{i<j}^NV^{\text{2B}}(i,j) 
	+ \sum_{i<j<k}^N V^{\text{3B}}(i,j,k) + \dots + V^{\text{NB}}(1,\dots,N).
\end{equation}
where $V^{\text{1B}}(i) = E(i) - E_{\text{eq}}(i)$ corresponds to
the one-body (1B) energy required to deform an individual water molecule (monomer) 
from its equilibrium geometry. 
All higher-order terms $V^{\text{nB}}$ in Eq.~\ref{eq:mbexpansion}
describe $n$-body (nB) interactions defined recursively as
\begin{equation} \label{eq:nbody}
	\begin{split}
	V^{\text{nB}}(1,\dots,n) = \quad & E_n(1,\dots,n) - 
	\sum_iV^{\text{1B}}(i) - \sum_{i<j}V^{\text{2B}}(i,j) - \dots \\
	& - \sum_{i<j<\dots<n-1}V^{\text{(n-1)B}}(i,j,\dots,(n-1)).
	\end{split}
\end{equation}

\noindent 
Most popular molecular models of water are pairwise additive (i.e., they truncate Eq.~\ref{eq:mbexpansion} 
at the 2B term) and use an effective $V^{\text{2B}}$ to account for many-body contributions 
in an empirical fashion.\cite{berendsen_hjc_1981, jorgensen_wl_1983, berendsen_hjc_1987,
dang_lx_1987, ferguson_dm_1995, mahoney_mw_2000, horn_hw_2004, wu_y_2006, paesani_f_2006, 
habershon_s_2009, park_k_2011, joung_is_2008}
Although in the early times of computer simulations this simplification was a necessity 
dictated by computational efficiency, 
the importance of many-body effects in water
was already recognized in the 1950s by Frank and Wen who introduced
a molecular model of liquid water consisting of ``flickering clusters of hydrogen-bonded molecules",
emphasizing the ``co-operative nature" of hydrogen bonding.\cite{frank_hs_1957}
It also became soon apparent that ``pair potentials do not realistically reproduce both gas and condensed 
phase water properties".\cite{lybrand_tp_1985}
The first attempts to derive potential energy functions (PEFs) for aqueous systems which could rigorously represent 
the individual terms of the MBE were made in the late 1970s and 1980s. 
\cite{matsuoka_o_1976, lie_gc_1986, lyubartsev_ap_2000, honda_k_1987, 
niesar_u_1989, niesar_u_1990}
In particular, Clementi and coworkers developed a series of analytical PEFs for water which
were fitted to \textit{ab initio} reference data obtained at the fourth-order
M{\o}ller-Plesset (MP4) and Hartree-Fock levels of theory for the 2B and 3B terms, respectively,
and represented many-body effects through a classical polarization term.
\cite{niesar_u_1989, niesar_u_1990}
Stillinger and David developed a polarizable model for water in which
H$^+$ and O$^{2-}$ moieties were considered as the basic dynamical and structural elements.
\cite{stillinger_fh_1978}
Building upon these pioneering studies, several polarizable models 
have been proposed over the years,
most notably the Dang-Chang model,\cite{dang_lx_1997}
the TTM models\cite{burnham_cj_2002, xantheas_ss_2002, burnham_cj_2002a, 
burnham_cj_2002b, fanourgakis_gs_2008, burnham_cj_2008}, and AMOEBA.\cite{ren_p_2003, wang_lp_2013}
The interested reader is referred to Refs.~\citenum{shvab_i_2016, cisneros_ga_2016} for recent reviews. 
Finally, in recent years, also machine learning potentials have been applied to water \cite{GaborPRB,PNASBehler}, which are able to include high order many body terms in the PEFs in form of structural descriptions of the atomic environments.

The development of efficient algorithms for correlated electronic structure methods along with
continued improvements in computer performance has recently made it possible to evaluate 
the individual terms of Eq.~\ref{eq:mbexpansion}, with chemical accuracy.
In parallel, tremendous progress has been made in constructing multidimensional mathematical functions that
are capable to reproduce interaction energies in generic N-molecule systems, with high fidelity.
\cite{braams_bj_2009, behler_j_2017, bartok2013machine}

By combining these three approaches, it has been realized that  
the MBE  provides a rigorous and efficient framework for the development of full-dimensional PEFs 
entirely from first principles, in which low-order terms are accurately determined 
from correlated electronic structure data, 
e.g., using coupled cluster theory with single, double, and perturbative triple excitations, CCSD(T), in the
complete basis set, CBS, limit, the current ``gold standard" for chemical accuracy, 
and higher-order terms are represented by classical many-body induction. 
Along these lines, several many-body PEFs for water have been proposed in the last decade, 
the most notable of which are CC-pol,\cite{bukowski_r_2007} 
WHBB,\cite{wang_y_2011} HBB2-pol,\cite{babin_v_2012} and 
MB-pol.\cite{babin_v_2013, babin_v_2014, medders_gr_2014}
When employed in computer simulations that allow for explicit treatment of nuclear quantum effects, 
these many-body PEFs have been shown to correctly predict structural, thermodynamic, dynamical,
and spectroscopic properties of water, from the dimer in the gas phase to liquid water and ice
(see Ref.~\citenum{cisneros_ga_2016} for a recent review).

Among the existing many-body PEFs, MB-pol (PIP-MB-pol in the present nomenclature) has been shown to correctly predict
the properties of water across different phases,\cite{paesani_f_2016} 
reproducing the vibration-rotation tunneling spectrum of the water dimer,\cite{babin_v_2013}
the energetics, quantum equilibria, and infrared spectra of small clusters,
\cite{babin_v_2014, richardson_jo_2016, cole_wts_2016, brown_se_2017}
the structural, thermodynamic, and dynamical properties of liquid water,\cite{medders_gr_2013, reddy_sk_2016}
including subtle quantum effects such as equilibrium isotope fractionation~\cite{chen+16jpcl},
the energetics of the ice phases,\cite{pham_ch_2017} the infrared and Raman spectra 
of liquid water,\cite{medders_gr_2015, straight_sc_2016} the sum-frequency generation spectrum 
of the air/water interface at ambient conditions,\cite{medders_gr_2016}
and the infrared and Raman spectra of ice I$_h$.\cite{moberg_dr_2017} 
It has been shown that the accuracy of PIP-MB-pol in reproducing the properties of water
depends primarily on its ability to correctly represent each individual term of the MBE 
at both short- and long-range.  

Briefly, within MB-pol, $V^{\text{1B}}$ in Eq.~\ref{eq:mbexpansion} is represented by the 1B PEF
developed by Partridge and Schwenke\cite{partridge_1997}, which reproduces intramolecular distortion 
with spectroscopic accuracy.
$V^{\text{2B}}$ includes a term describing 2B dispersion, which is derived from the asymptotic expansion 
of the interaction energy, as well as a term 
describing electrostatic interactions associated with both permanent and induced molecular moments.
At short-range, within the original PIP-MB-pol, $V^{\text{2B}}$ is supplemented by a 4$^{th}$-degree permutationally invariant polynomial 
(PIP)\cite{braams_bj_2009}
that smoothly  switches to zero as the distance between the two oxygen atoms 
in the dimer approaches 6.5 \AA.\cite{babin_v_2013}
Similarly, $V^{\text{3B}}$ includes a 3B induction term that is supplemented by a short-range 4$^{th}$-degree 
PIP that smoothly switches to zero
once the oxygen-oxygen distance between two pairs of water molecules within the trimer 
approaches 4.5 \AA.\cite{babin_v_2014}
All higher-body terms are implicitly represented by classical many-body induction according
to a modified Thole-type scheme originally adopted by the TTM4-F water model.\cite{burnham_cj_2008}
The PIP 2B and 3B terms, which were derived from CCSD(T) calculations carried out in the complete
basis set limit for large sets of water dimers and trimers, correct for deficiencies associated
with a purely classical description of intermolecular interactions by effectively 
representing quantum-mechanical interactions that arise from the overlap of the monomer electron 
densities (e.g., charge transfer and penetration, and Pauli repulsion).

In this study, we investigate the application of Behler-Parrinello neural networks\cite{behler_j_2007, behler_j_2015} (BPNN) and 
Gaussian approximation potentials (GAP) as alternatives for the original PIP representations 
of MB-pol short-range 2B and 3B terms.
Using the same training, validation, and test sets,
two additional (BPNN- and GAP-based) analytical expressions of MB-pol are derived,
which effectively exhibit the same accuracy as the original, PIP-based expression.
This study provides further evidence for the ability of the MBE in combination with machine learning techniques to	serve as a rigorous and efficient route for	the development	of accurate potential energy functions such as MB-pol in the	case of	water.
The article is organized as follows: In Section~\ref{sect:mbpol_comp}, we provide
an overview of the computational framework associated with the many-body formalism adopted 
by MB-pol, while in Section~\ref{sect:models} we describe the three different models
(PIP-MB-pol, BPNN-MB-pol, and GAP-MB-pol) used to represent water two-body and three-body interactions.
The results are presented in Section~\ref{sect:results}, and the conclusions 
along with an outlook are given in Section~\ref{sect:conclusions}.

\section{MB-pol functional form and computational details} \label{sect:mbpol_comp}

We are employing the MB-pol framework for water, which is based on the MBE of Eq.\ \eqref{eq:mbexpansion} and contains explicit terms for the 1B, 2B, and 3B terms, in combination with classical N-body polarization that accounts for all higher-body contributions to the interaction energy\cite{babin_v_2013,babin_v_2014}.
In MB-pol, the 2B term is divided into long-range interactions that are well described using classical expressions for electrostatics, induction, and dispersion, and short-range interactions that include complex quantum-mechanical effects due to the overlap of the monomer electron densities.
\begin{equation}
  V^{\text{2B}}(i,j)
  =V^{\text{2B}}_{\text{short}}(i,j)+V^{\text{2B}}_{\text{long}}(i,j)
\end{equation}
with
\begin{equation}
  V^{\text{2B}}_{\text{long}}(i,j)
  =V^{\text{2B}}_{\text{TTM,elec}}(i,j)
  +V^{\text{2B}}_{\text{TTM,ind}}(i,j)
  +V^{\text{2B}}_{\text{disp}}(i,j),
\end{equation}
where $V^{\text{2B}}_{\text{TTM,elec}}$ and $V^{\text{2B}}_{\text{TTM,ind}}$ are electrostatic and induction energies represented by a slightly modified version of the Thole-type TTM4-F model\cite{burnham_cj_2008,babin_v_2013,babin_v_2014}, and the dispersion energy $V^{\text{2B}}_{\text{disp}}$ is modeled by a $C_6$ term that is dampened at short range\cite{babin_v_2013}.
Similarly, the 3B term in MB-pol is decomposed into classical 3B induction that captures essentially all of the 3B interaction energy at long range and an expression for the highly complex interactions at short range,
\begin{equation}
  V^{\text{3B}}(i,j,k)
  =V^{\text{3B}}_{\text{short}}(i,j,k)+V^{\text{3B}}_{\text{TTM,ind}}(i,j,k).
\end{equation}
Because corrections to the underlying classical baseline potentials $V^{\text{2B}}_{\text{long}}$ and $V^{\text{3B}}_{\text{long}}=V^{\text{3B}}_{\text{TTM,ind}}$ are only required at short range, and in order to obtain a smooth, differentiable potential energy surface, MB-pol employs switching functions that smoothly turn off the short-range potentials $V^{\text{2B}}_{\text{short}}$ and $V^{\text{3B}}_{\text{short}}$ once the separation between the oxygen atoms of the water molecules exceeds a preset cutoff.

The MB-pol short-range 2B and 3B potentials\cite{babin_v_2013,babin_v_2014} are written as
\begin{equation}
  V^{\text{2B}}_{\text{short}}(i,j)
  =s(i,j)V^{\text{2B}}_{\text{ML}}(i,j)
  \label{fswitch2B}
\end{equation}
and
\begin{equation}
  V^{\text{3B}}_{\text{short}}(i,j,k)
  =s(i,j,k)V^{\text{3B}}_{\text{ML}}(i,j,k),
\label{fswitch3B}
\end{equation}
where
\begin{equation}
  s(i,j,k)
  =s(i,j)s(i,k)+s(i,j)s(j,k)+s(i,k)s(j,k).
\end{equation}
The switching function was chosen as
\begin{equation}
  s(i,j)=
  \begin{cases}
    1                                 & \text{if } t_{ij}<0\\
    \cos^2\left(\frac{\pi}{2}t_{ij}\right) & \text{if } 0\leq t_{ij} < 1\\
    0                                 & \text{if } 1 \leq t_{ij}
  \end{cases},
  \label{fswitch-eq}
\end{equation}
where 
\begin{equation}
  t_{ij}=\frac{R^{\text{OO}}_{ij}-R_{\text{low}}}
              {R_{\text{high}}-R_{\text{low}}}
\end{equation}
is a scaled and shifted oxygen-oxygen distance for water molecules $i$ and $j$.
The MB-pol 2B and 3B cutoff values are $R_{\text{low}}^{\text{2B}}=4.5$\,\AA, $R_{\text{high}}^{\text{2B}}=6.5$\,\AA, $R_{\text{low}}^{\text{3B}}=0.0$\,\AA, and $R_{\text{high}}^{\text{3B}}=4.5$\,\AA.

An accurate description of both the 2B and the 3B short-range interactions requires flexible multi-dimensional functions, for which the original PIP-MB-pol model employs permutationally invariant polynomials\cite{braams_bj_2009,xie_z_2010} (PIPs).
In this work we investigate the performance of alternative machine learning (ML) frameworks to represent these 2B and 3B short range interactions in water, by comparing PIPs to Behler-Parrinello neural networks (BPNN) and Gaussian approximation potentials (GAP) for $V^{\text{2B}}_{\text{ML}}$ and $V^{\text{3B}}_{\text{ML}}$.
We employ the original MB-pol switching functions and cutoff values with the PIP and BPNN potentials, while GAP uses slightly different cutoff values and switching functions.\cite{bartok2015gaussian}
In the context of MBE and neural networks, it should be noted that a neural network representation of the many-body expansion of the interaction energy, truncated at the 3B term, has been reported for methanol.\cite{yao_k_2017}

\subsection{Training sets and reference energies} \label{sect:E_ref}

We employ the original MB-pol 2B and 3B data sets\cite{babin_v_2013,babin_v_2014}, which sample regions of the 2B and 3B water PES, respectively, that are most relevant for simulations of water at normal to moderate temperature and pressure.
The 2B training set consists of 42,508 water dimer structures with center-of-mass separations ranging from 1.6 to 8\,\AA\ that include the global dimer minimum geometry, several saddle points, compressed geometries with positive interaction energies, and dimers extracted from path-integral molecular dynamics (PIMD) simulations of liquid water at ambient temperature and pressure.
Similarly, the 3B training set contains 12,347 water trimer structures that include the global minimum and trimers extracted from a range of MD and PIMD simulations of small water clusters, liquid water, and water ice phases at varying temperatures and pressures.
Both the 2B and the 3B QM reference energies of these data sets were obtained at the complete basis set (CBS) limit of coupled cluster theory with single, double and iterative triple excitations, CCSD(T). For details see the original publications\cite{babin_v_2013,babin_v_2014}.
The short-range training set energies $V_{\text{short}}^{\rm ref}$ employed in this work were obtained from the QM reference data by subtracting the MB-pol baseline long-range 2B and 3B potentials $V^{\text{2B}}_{\text{long}}$ and $V^{\text{3B}}_{\text{TTM,ind}}$, respectively.


The original 2B dataset includes a few dimer structures with extremely high binding energy. Those high energy structures are not only physically unimportant, but also sparsely distributed, which can lead to difficulties for machine learning techniques to make effective predictions for structures in this regime because of insufficient information. Therefore, we have retained only configurations with binding energies below 60\,kcal/mol in this work. In addition, we have removed all configurations with oxygen-oxygen separations larger than the MB-pol 2B short-range cutoff of 6.5\,\AA, leading to a total of 42,069 configurations in the final 2B training set.
In contrast, the trimer dataset with 12,347 configurations is fully employed.
Each dataset is then randomly divided into three separate sets, training, validation, and test sets with a ratio of 0.81:0.09:0.1. The first two are used during training and for model selection while the last one is kept completely isolated from the training procedure and is employed for the final evaluation only.

\subsection{Water cluster test sets}

Reference interaction energies of (H$_2$O)$_n$ clusters with $n=4-6$ (see Fig.\ \ref{figure-4}) are based on geometries optimized with MP2 and RI-MP2\cite{bates_dm_2009,temelso_b_2011} and were taken from Ref.\ \citenum{reddy_sk_2016}.
The energies were obtained using the MBE of the interaction energy\cite{gora_u_2011} with both 2B and 3B interaction energies computed at the same level as the MB-pol 2B and 3B training sets\cite{babin_v_2013,babin_v_2014}, that is, effectively at the CBS limit of CCSD(T).
All higher order contributions to the interaction energy ($>$3B) were obtained from explicitly correlated CCSD(T)-F12b\cite{adler_tb_2007} calculations with the VTZ-F12 basis set\cite{peterson_ka_2008}, which yields results close to the CBS.

\section{Many-body models} \label{sect:models}
\subsection{Permutationally invariant polynomials}

The permutationally invariant polynomials are functions of the distances between pairs involving both the physical atoms (H and O) and two additional sites L$_1$ and L$_2$ that are located symmetrically along the directions of the oxygen lone pairs of a water molecule,
\begin{equation}
  \bvr_{\text L}^{(\pm)}
  =\bvr_{\text{O}}
  +\frac{1}{2}\gamma_{\parallel}(\bvr_{\text{OH}_1}+\bvr_{\text{OH}_2})
  \pm\gamma_{\perp}(\bvr_{\text{OH}_1}\times\bvr_{\text{OH}_2}),
\end{equation}
where $\gamma_{\parallel}$ and $\gamma_{\perp}$ are fitting parameters and $\bvr_{\text{OH}_{1,2}}$ are the O-H bond vectors.
Exponential functions of the type $\xi_i=e^{-kd_i}$ or $\xi_i=e^{-k(d_i-d^{(0)})}$ and Coulomb-type functions $\xi_i^{\text{Coul}}=e^{-kd_i}/d_i$ are built for the set $\{d_i\}$ of these distances and are used as basis for the PIPs.
The PIP $V_{\text{ML,PIP}}=\sum_lc_l\eta_l$ is then constructed from the set $\{\xi_i\}$ of these functions, where $\{\eta_l\}$ are symmetrized monomials up to a given degree. The symmetrization is carried out such that the monomials, and hence the PIP, are invariant with respect to the permutations of the water molecules as well as to the permutations of equivalent H and L sites within each molecule.
The polynomial coefficients $c_l$ and the exponential coefficients $k$ and distances $d^{(0)}$ are linear and non-linear fitting parameters, respectively.
For details see the original publications\cite{babin_v_2013,babin_v_2014}. 

For the 2B PIP we are using 31 basis functions:
6 exponential functions for all intra-molecular HH and OH pairs with exponential coefficients
$k_{\text{HH}}^{\text{intra}}$ and
$k_{\text{OH}}^{\text{intra}}$;
9 Coulomb-type functions for all inter-molecular HH, OH and OO pairs with exponential coefficients
$k_{\text{HH}}^{\text{inter}}$,
$k_{\text{OH}}^{\text{inter}}$, and
$k_{\text{OO}}^{\text{inter}}$;
15 exponential functions for all inter-molecular LH, LO and LL pairs with exponential coefficients
$k_{\text{LH}}^{\text{inter}}$,
$k_{\text{LO}}^{\text{inter}}$, and
$k_{\text{LL}}^{\text{inter}}$.
A total of 1153 symmetrized monomials form $V^{\text{2B}}_{\text{ML,PIP}}$: 
6 first-degree monomials using only intermolecular $\xi_i$ variables, 
63 second-degree monomials with at most a linear dependence on intramolecular variables, 
491 third-degree ones containing at most quadratic intramolecular variables, 
593 fourth-degree terms involving only quadratic intramolecular variables, as in the original paper \cite{babin_v_2013}. 

For the 3B PIP we are using 36 exponential functions for each of the intra- and inter-molecular distances between all real (O and H) atoms with exponential coefficients and distances
$k_{\text{HH}}^{\text{intra}}$,
$k_{\text{OH}}^{\text{intra}}$,
$k_{\text{HH}}^{\text{inter}}$,
$k_{\text{OH}}^{\text{inter}}$,
$k_{\text{OO}}^{\text{inter}}$,
$d_{\text{HH}}^{\text{intra},(0)}$,
$d_{\text{OH}}^{\text{intra},(0)}$,
$d_{\text{HH}}^{\text{inter},(0)}$,
$d_{\text{OH}}^{\text{inter},(0)}$, and
$d_{\text{OO}}^{\text{inter},(0)}$.
A total of 1163 symmetrized monomials form $V^{\text{3B}}_{\text{ML,PIP}}$:  
13 second-degree monomials with only intermolecular exponential variables, 
202 third-degree monomials with at most a linear dependence on intramolecular variables, 
948 fourth-degree monomials containing at most a linear dependence on intramolecular variables 
or intermolecular ones involving oxygen-oxygen and hydrogen-hydrogen distances,
as in the original paper \cite{babin_v_2014}.

The linear and nonlinear parameters were optimized using a singular value decomposition and the simplex algorithm, respectively, by minimizing the regularized sum of squared errors $\chi^2$ for the corresponding training set $S$, commonly referred to as Tikhonov regularization or ridge regression\cite{tikhonov_an_1963},
\begin{equation}
  \chi^2
  =\sum_{n\in S}[V_{\rm short}(n)-V_{\text{short}}^{\text{ref}}(n)]^2
  +\Gamma^2\sum_lc_l^2.
\end{equation}
The regularization parameter $\Gamma$ was set to 5$\times 10^{-4}$ for 2B and 1$\times 10^{-4}$ for 3B in order to reduce the variation of the linear parameters without spoiling the overall accuracy of the fits.

\subsection{Behler-Parrinello neural networks}

Based on the assumption that the total energy of a system can be written as a sum of atomic energy contributions, a BPNN consists of a set of fully connected feed-forward neural networks, each of which provides an atomic energy\cite{behler_j_2007, behler_j_2015}. 
Each atomic network takes as its input a set of atom-centered symmetry functions\cite{atomcentered} 
that encode the atomic positions and at the same time are invariant with respect to overall rotation and translation, as well as to permutations of like atoms. 
The invariance of the total energy is assured by enforcing that all atomic networks of the same species are identical, thus having the same structure and weights. 
As a result, for the water systems considered here, there are two sub-networks, one for all H atoms and the other for all O atoms, which need to be trained simultaneously. 
Aiming at smoothly disabling the short-range interaction energy contribution at long distances, described in Eqs. \ref{fswitch2B}-\ref{fswitch3B}, 
the sum of all atomic energies from the last layer of the sub-networks is multiplied by the switching function to produce a final output for a BPNN. 
The network weights are determined with respect to the values of the reference short-range interaction energies.    

The following modified radial and angular symmetry functions, which lack the cut-off functions of the original BPNN approach, have been chosen for each atom $i$
\be
G_{i}^{rad} = \sum_{j \neq i} e^{-
\eta (R_{ij}-R_s)^2}, 
\label{G_rad} 
\ee
\be
G_{i}^{ang} = 2^{1-\zeta}\sum_{j \neq i} \sum_{k \neq i, j}  
(1+\lambda \cos \theta_{ijk})^\zeta
e^{-\eta' (R_{ij} + R_{ik} + R_{jk})^2}, 
\label{G_ang}
\ee
resulting in an input vector ${\bf G}_i = \{G_{i}^{rad/ang} \}$ for the atomic network. 
$\theta_{ijk}$  denotes the angle enclosed by two interatomic distances $R_{ij}$ and $R_{ik}$. Each summation above takes into account only same combination of atomic species and the set of parameters, $\{(\eta, R_ s)\}$ and $\{(\zeta, \lambda, \eta')\}$, is the same for each type of species grouping. We have removed the cut-off function from the original forms of the symmetry functions used in Ref. \onlinecite{behler_j_2007,behler_j_2015} since we apply the MB-pol 2B and 3B switching functions, thus never feeding any structures to the 2B and 3B BPNNs that are beyond the cut-off region.

The dimension of the input vector should reflect a balance between giving an effective resolution of the local environment and the computational cost of training and inference with a large input vector neural network. 
After carefully examining different parameter sets, we have come up with the final set as follows. 
For the 2B term, there are 24 radial Gaussian-shape filters, Eq.\ \eqref{G_rad},
whose centers $R_s$ are placed evenly between 0.8\,\AA\ and 8\,\AA, which are relatively close to the smallest and the largest interatomic distances in the training set.
For O-O distances the two smallest centers are excluded because the O-O separation is well beyond the space covered by these two filters. 
The width of those filters is proportional to their centers' position, $1/\sqrt{2\eta} = 0.2R_s$. 
The angular probe in Eq. \ref{G_ang} takes $\zeta=[1, 4, 16]$ for different filter widths, $\lambda=\pm 1$ for switching the filter's center between 0 and $\mathrm{\pi}$, and  $\eta'=\left[ 0.001, 0.01, 0.05 \right]$ ($\mathrm{\mathring{A}^{-2}}$) for various levels of the separation dependence. 
As for 3B BPNN, a similar scheme is applied with few adjustments,
 which include 16 radial filters with centers arranged in the same range, between 0.8\,\AA\ and 8\,\AA, 
 and two levels of separation dependence attached to the angular filter, $\eta'=\left[ 0.001, 0.03 \right]$ ($\mathrm{\mathring{A}^{-2}}$). 
 Moreover, to reduce the redundancy and computational cost, for the angular probe for hydrogen atoms, 
 we consider only two types of triplet of atoms, a hydrogen atom with other two hydrogen atoms or with an oxygen and another hydrogen. 
In total, a set of 82 and 84 symmetry functions for O and H is formed for the 2B BPNN 
while another set of 66 and 56 functions for O and H is used for the 3B BPNN.
The complete set of the symmetry function parameters can be found in the SI.

 The neural network training encounters various hyperparameters and different techniques for initialization of these parameters, which are mostly found by trial and error. Following is our final network architecture and set-up for the network training. 
The atomic network consists of one input layer, three hidden layers, and one single output layer. 
The input layer takes as its input the preprocessed symmetry functions, each of which is obtained by rescaling the symmetry function with its corresponding maximum value in the training and validation sets. 
Furthermore, the numbers of units in each hidden layer are chosen to be the same for both atomic networks for O and H.
Overall, with 34 and 22 units per hidden layer, the final 2B and 3B BPNN models contain 10542 and 4798 weight and bias parameters, respectively.
 For the continuity of the energy functional, the activation function for each unit is chosen to be a hyperbolic tangent for the hidden layers and a linear function for the output layer. 
Besides, the reference energies for the 2B training are converted to energy per atom in eV unit so that the network targets a similar range of values as given by the activation functions. 

We build the network models using Keras\cite{chollet2015keras}
with Theano\cite{2016arXiv160502688short}
backend and choose the Adam optimizer with a batch size of 64 for training. The Nguyen-Widrow method \cite{nguyen_d_1990} is employed to initialize the network weights and biases. 
For a stable and effective training, the optimization process is continuously carried out five times with descending starting learning rates $\left[ 10^{-3}, 2\cdot 10^{-4}, 6 \cdot 10^{-5}, 9 \cdot 10^{-6}, 10^{-6}\right]$  and corresponding numbers of iterations, or epochs, $\left[1500, 1500, 1000, 1000, 1000\right]$. 
Furthermore, we apply an additional decay rate $\alpha=10^{-5}$ to each learning rate such that at a given epoch $k$ the leaning rate is $lr_{k-1}/(1+\alpha \cdot k)$ based on the value at the previous epoch $lr_{k-1}$. 
The training is to optimize the mean squared error of the modeled energies compared to the reference data in the training set. 
To avoid overfitting, on each epoch, the quality of the model is monitored on the validation set such that only the model that gives the highest accuracy over this set is ultimately kept. 
Finally, the trained model is then evaluated on the test set to quantify its capability of generalization to unseen data. 
For the systems considered here, the training processes generally take three hours and one hour on a Tesla K40 GPU with the GPU-accelerated cuDNN library for 2B and 3B sets, respectively.

\subsection{Gaussian Approximation Potentials}

The Gaussian Approximation Potential (GAP)\cite{bartok_ap_2010,bart+13prb} framework, available in the QUIP program package\cite{QUIPwebpage}, is an implementation of Gaussian process regression (GPR) interpolation for the atomic energy as a function of the  geometry of the neighbouring atoms. The functional form representing a function $f$ that is to be interpolated is identical to that of kernel ridge regression, 
\begin{equation}
    f({\bm R}) = \sum_k b_k K({\bm R},{\bm R}_k),
\end{equation}
where the high dimensional vector ${\bm R}$ represents the complete geometry of neighbouring atoms, $k$ indexes a set of representative data points $\{{\bm R}_k\}$, $K$ is the kernel function, and $\{b_k\}$ are fitting coefficients. In the GPR formalism, $K$ corresponds to an estimate of the covariance of the unknown function, and the linear system is solved in the least squares sense using Tikhonov regularisation, but the regularisation parameters are now interpreted as estimates of data and model error. In the present case, the regularisation was chosen to be 0.00115\,kcal/mol for the 2B term, and 0.0231\,kcal/mol for the 3B term after manual exploration of the data. 

The success of the GAP fit depends on choosing an appropriate kernel, one that captures the structure of the input data and as much as possible about the function to be fitted. Here we use the "Smooth Overlap of Atomic Positions" (SOAP), a kernel that is the rotationally integrated overlap of the neighbour densities, which was shown to be equivalent to the scalar product of the spherical Fourier spectrum\cite{bart+13prb}. The atomic environment of atom $i$ is described by a set of neighbour densities, one for each atomic species, which are represented as the sum of Gaussians each centred on one of the neighbouring atoms $j$ \cite{de+16pccp}:
\begin{equation}
   \rho_{i}^\alpha ({\bm r}) = \sum_{j}{ \exp {\bigg(-  \frac{|{\bm r}-{\bm r}_{ij}|^2}{2 \sigma_{at}^2}\bigg) }} f_{cut}( {\bm r}_{ij})
\end{equation}
where $j$ ranges over neighbours with  atomic species $\alpha$, ${\bm r}_{ij}$ are the positions relative to $i$, and $\sigma_{at}^2$ is a smoothing parameter. We included the switching function $f_{cut}$ which smoothly goes to zero beyond a specified radial value. This local atomic neighbour density can be expanded in terms of spherical harmonics, $Y_{lm}(\hat{\bm r})$ and orthogonal radial functions, $g_n(|\bm r|)$ :
\begin{equation}
    \rho_{i}^\alpha ({\bm r})= \sum_{nlm} c_{nlm}^\alpha g_n (| \bm r|) Y_{lm}(\hat{{\bm r}})
\end{equation}
The expansion coefficients are then combined to form the rotationally invariant power spectrum:
\begin{equation}
    p_{n_1 n_2 l}^{\alpha\beta}({\bm R}_i) = \pi \sqrt{\frac{8}{2l+1}} \sum_m (c^\alpha_{n_1lm})^{\dagger} (c^\beta_{n_2lm})
\end{equation}
where we have emphasized the functional dependence on the complete neighbour geometry. The complete SOAP kernel can be written as:
\begin{equation}
    K({\boldsymbol{R},\boldsymbol{R'}}) = \bigg( \sum_{\alpha\beta n_1 n_2 l} p^{\alpha\beta}_{n_1 n_2 l}({\boldsymbol{R}}) p^{\alpha\beta}_{n_1 n_2 l}({\boldsymbol{R'}}) \bigg)^\zeta,
\end{equation}
where we have allowed for a small integer exponent $\zeta$ (here set to 2). The kernel is  also normalised so that the kernel of each environment with itself is unity. Separate fits are
made to the atomic energy function corresponding to each atomic species taken as the center of an atomic environment. The key free parameters are the radial cutoff in $f_{cut}$, and the smoothing parameter $\sigma_{at}$. In the present cases here, atomic energy functions are represented by the sum of two kernels\cite{bart+17sa}, one with a smaller radial cutoff (4.5\,\AA) and smaller smoothing (0.4\,\AA), and one with a larger cutoff (6.5\,\AA\ for the 2B and 7.0\,\AA\ for the 3B fit) and larger smoothing (1.0\,\AA). The RMSE is only weakly sensitive to these, and some manual optimisation was carried out.  Each fit uses 10 radial basis functions and a spherical harmonics basis band limit of 10. The representative environments for the fit are chosen using CUR matrix decomposition\cite{maho-drin09pnas}. The number of representative points are 9000 in the 2B fit and 10000 in the 3B fit. The full command lines of the fits are given in the Supporting Information. Note that although formally the GAP construction corresponds to a decomposition of the total energy into atomic energies, similarly to BPNN above, the cutoffs are sufficiently large to encompass all atoms in the water dimer and trimers in the dataset, and therefore the decomposition does not represent an approximation.

\section{Results} \label{sect:results}

\subsection{2B and 3B interactions, and the structure of the training data}

\label{sect:mb_interactions}
The root mean squared errors (RMSEs) obtained with PIPs, BPNNs, and GAPs for
the 2B and 3B datasets are reported in Table \ref{rmse-2b-3b-table}. 
\begin{table}[h]
  \centering
\caption{RMSE (in kcal/mol) per isomer on the provided training, validation, and test sets in the PIP, BPNN, GAP short
range interaction two-body (2B) and three-body (3B) energy fitting.}
\label{rmse-2b-3b-table}
\begin{tabular}{c c ccc c ccc}
\noalign{\smallskip}\hline 
  &  &  \multicolumn{3}{c}{2B}  &  &   \multicolumn{3}{c}{3B} \\  
 \cline{3-5}  \cline{7-9} 
  &    &  training  &  validation  &  test  &  &    training  &  validation  &  test \\
\noalign{\smallskip}\hline
PIP   &   &  0.0349  &  0.0449  &  0.0494   &  &    0.0262  &  0.0463  &  0.0465 \\
BPNN  &    &  0.0493  &  0.0784  &  0.0792  &  &    0.0318  &  0.0658  &  0.0634 \\
GAP  &    &  0.0176  &  0.0441  &  0.0539   &  &    0.0052  &  0.0514  &  0.0517\\
\noalign{\smallskip}\hline
\end{tabular}
\end{table}
For the 2B term, all three methods achieve similar accuracy: 
the error on the training set is less than 0.050 kcal/mol per dimer 
while the errors on validation and test sets are less than 0.080 kcal/mol per dimer.
These errors demonstrate a high level of accuracy since the average value of the target energies 
in the dataset is 3 kcal/mol. 
Among the three, the 2B PIP model appears to perform better on the validation 
and test sets and suffers less from overfitting. 
The difference in RMSEs for the training set and the test set are below 0.02 kcal/mol with PIP, 
but around 0.03 kcal/mol with BPNN and 0.04 kcal/mol with GAP.
The GAP model gets a slightly lower error for the training set, but overfitting prevents to achieve a similar accuracy for the test set.   

\begin{figure*}[!ht]
\includegraphics[width=0.8\textwidth]{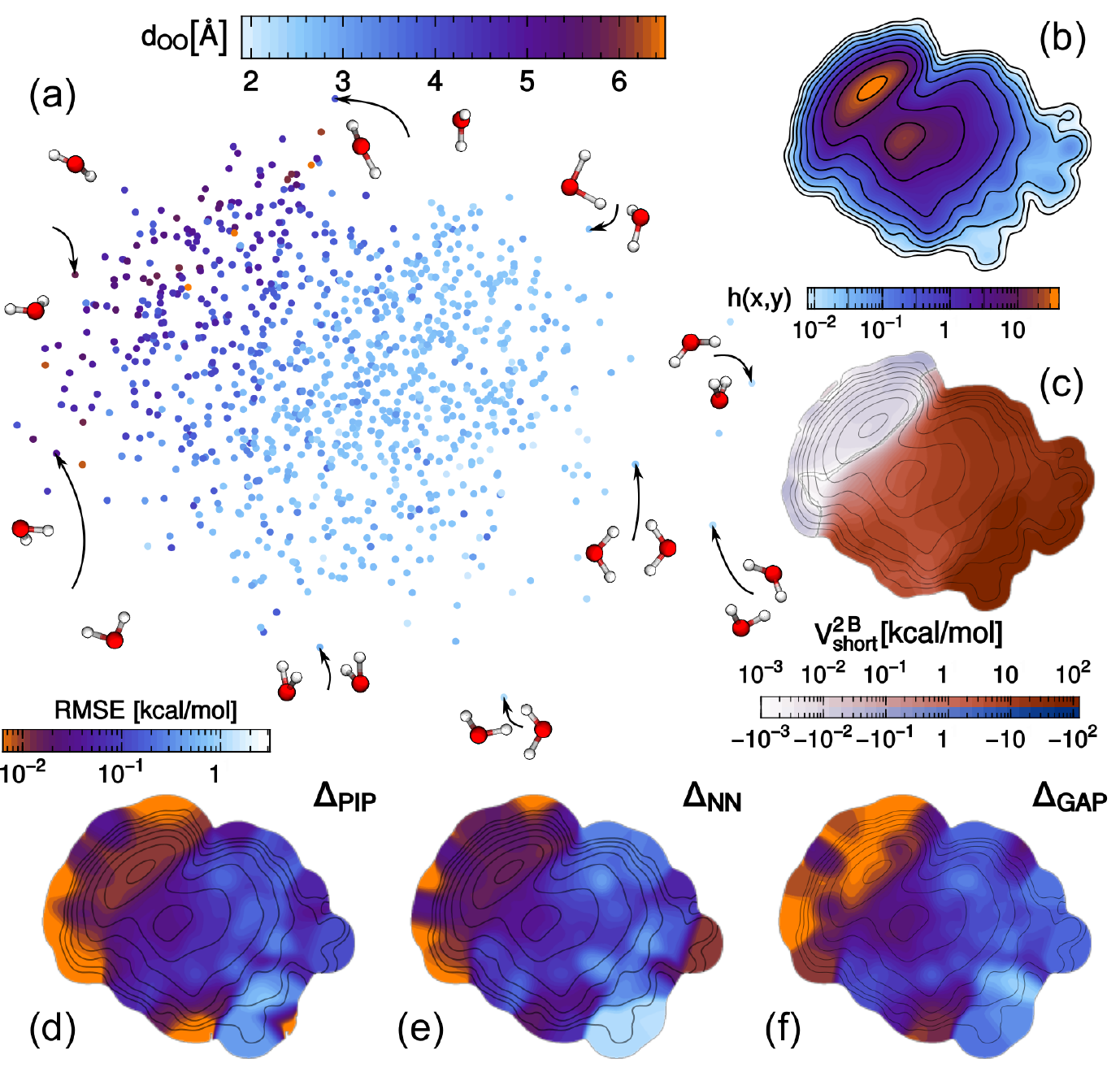}
\caption{(a) Sketch-map representation for the training data set for dimer configurations. Points are colored according to O-O distance, and a few reference configurations are also shown. (b) Histogram of the training point positions on the sketch-map. The train set density is also reported on other plots as a reference for comparison. (c) Conditional average of the 2B energies for different parts of the train set. (d-f) Conditional average RMSE for the PIP, BPNN, GAP fits of the 2B energy in different parts of the test set. }
\label{figure-1}
\end{figure*}

In order to investigate in more detail the performance of the different regression schemes for predicting the 2B and 3B energies over the MB-pol dimer and trimer data sets, we used a dimensionality reduction scheme to obtain a 2D representation of the structure of the train set. 
We followed a procedure similar to that used in Ref.~\citenum{de+16jci} to map a database of oligopeptide conformers. A metric based on SOAP descriptors~\cite{de+16pccp} was used to assess the similarity between reference conformations of dimers or trimers. A 2D map that best preserved the similarity between 1000 reference configurations selected by farthest point sampling~\cite{rose+77siam} was obtained using the sketch-map algorithm~\cite{ceri+11pnas,ceri+13jctc}. All other configurations (training and testing) were then assigned 2D coordinates $(x_i,y_i)$ by projecting them on the same reference sketch-map.
We could then compute the histogram of configurations $h(x,y)$, the averages of the properties of the different configurations, and of the test RMSE for the various methods, conditional on the position on the 2D map, e.g.
\begin{equation}
\begin{split}
    h(x,y) =& \left<\delta(x-x_i)\delta(y-y_i)\right>\\
    V_\text{short}^\text{2B}(x,y) = & \frac{\left<V_\text{short}^\text{2B}(i) \delta(x-x_i)\delta(y-y_i)\right>}{h(x,y)}.
    \end{split}    
\end{equation}

Figure~\ref{figure-1} demonstrates the application of this analysis to the dimer dataset.
One of the sketch-map coordinates correlates primarily with O-O distance, while different relative orientations and internal monomer deformations are mixed in the other direction.
Conformational space is very non-uniformly sampled (Fig.~\ref{figure-1}b), with a large number of configurations at large O-O distance -- which correspond to $V_\text{short}^\text{2B}$ of less than 0.01 kcal/mol -- and at intermediate distances, with sparser sampling in the high-energy, repulsive region (Fig.~\ref{figure-1}c). 
It is interesting to see that the three regression schemes we considered exhibit very similar performance in the various regions, with tiny errors $<0.01$ kcal/mol for far-away molecules, and much larger errors, as large as 1 kcal/mol, for configurations in the repulsive region. 
These large errors are not only due to the high energy scale of $V_\text{short}^\text{2B}$ in this region: the largest errors appear in the portion of the map which is characterized by both large $V_\text{short}^\text{2B}$ and low density of sample points.

\begin{figure}[!ht]
\includegraphics[width=0.45\textwidth]{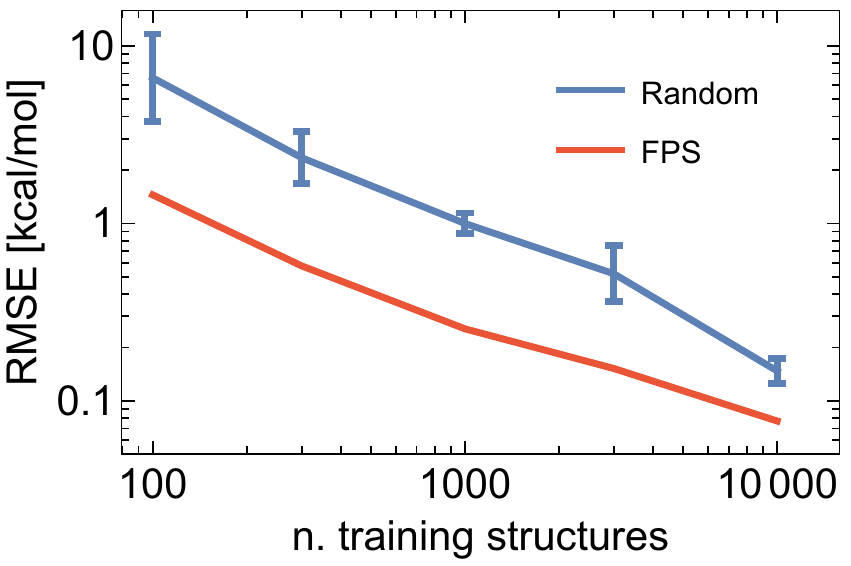}
\caption{TEST RMSE as a function of the size of the train set for the 2B energy contribution, using a BPNN for the regression. Training configurations were selected at random (5 independent selections, average and standard deviation shown) or by farthest point sampling. }
\label{figure-2}
\end{figure}

The non-uniform sampling of the dimer space configuration means that there is room to improve it. Figure~\ref{figure-2} compares the test RMSE obtained by BPNN fits constructed on subsets of the overall training set. 
The error can be reduced by up to a factor of five by choosing the subset with a FPS strategy, rather than at random. This observation is consistent with recent observations made using SOAP-GAP in a variety of systems~\cite{bart+17sa,musi+18cs}. 
Selecting training configurations from a larger database of potential candidates using FPS gives a viable strategy to reduce the number of high-end calculations that have to be performed to describe accurately interactions in the construction of a MB potential.

\begin{figure*}[!ht]
\includegraphics[width=0.8\textwidth]{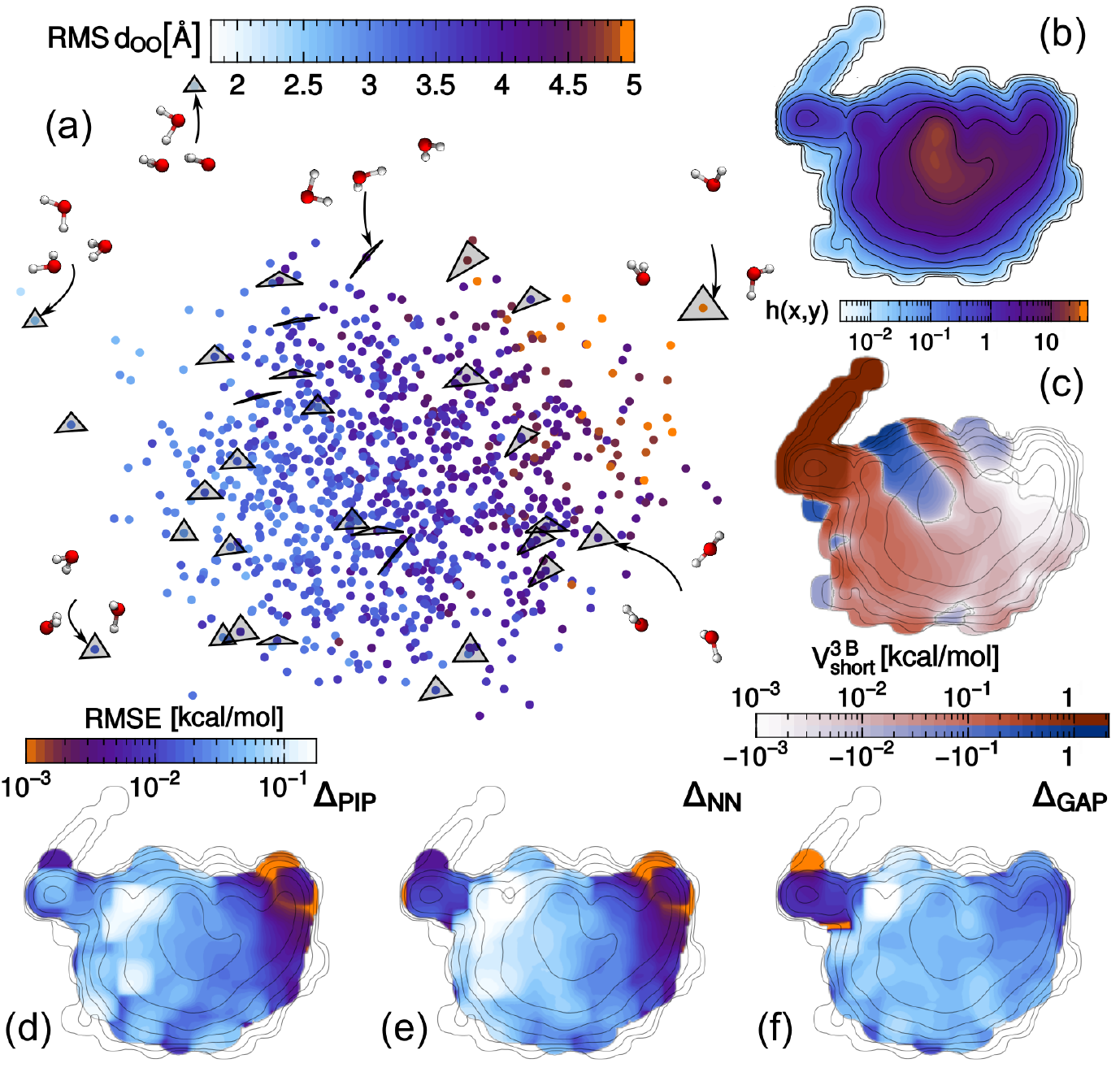}
\caption{(a) Sketch-map representation for the training data set for trimer configurations. Points are colored according to the root mean square of the three O-O distances; trimer geometries are also represented as triangles, together with a few structures for which a snapshot is shown. (b) Histogram of the training point positions on the sketch-map.  The train set density is also reported on other plots as a reference for comparison. (c) Conditional average of the 3B energies for different parts of the train set. (d-f) Conditional average RMSE for the PIP, BPNN, GAP fits of the 3B energy in different parts of the test set. }
\label{figure-3}
\end{figure*}

Figure~\ref{figure-3} shows a similar analysis for the case of the trimer data and $V^\text{3B}_\text{short}$. 3B energies span a smaller range than the 2B component, that includes most of the core repulsion. The higher dimensionality of the problem, however, makes this a harder regression problem, as is apparent from the irregular correlations between energy and position on the map, that reveals an alternation of regions of positive and negative contributions. 

As a result, the absolute RMSE accuracy of the regression models is comparable to that for the 2B terms, with PIP and GAP yielding comparable accuracy (RMSE $\approx 0.05$ kcal/mol), followed closely by BPNN (RMSE $\approx 0.06$ kcal/mol). As in the case of 2B energy contributions, an analysis of the error distribution shows that improving the sampling density and uniformity for the train set is likely to be the most effective strategy to further improve the model. Errors are concentrated at the periphery of the data set. The good performance of the GAP model can be traced to the fact that it provides a very good description of the short RMS $d_\text{OO}$ region, even if only a few reference structures are available, even though it performs less well than PIP or NN for configurations that involve far away molecules.

\subsection{Water clusters} \label{sect:clusters}

\begin{figure}[!ht]
  \includegraphics[width=0.7\textwidth]{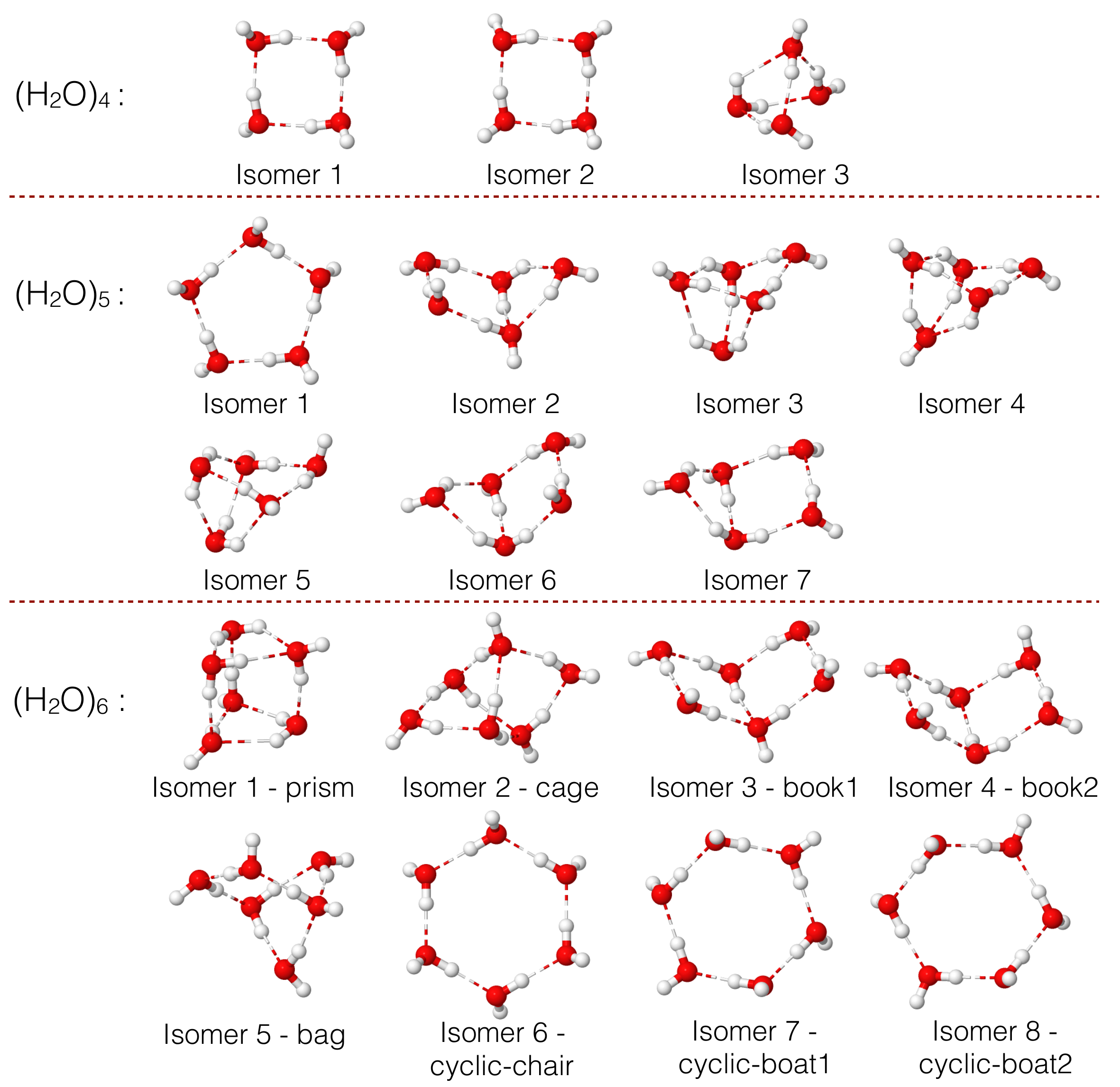}
  \caption{Isomers of water clusters (H$_2$O)$_n$, $n=4, 5, 6$, used for the analysis 
  of the performance of PIP, BPNN and GAP representations of 2B and 3B energies. Reproduced from Ref.\ \citenum{reddy_sk_2016}.}
  \label{figure-4}
\end{figure}

Isomers of water clusters (H$_2$O)$_n$ with $n=4, 5, 6$ (see Fig.\ \ref{figure-4} for the structures) 
serve as larger test systems to investigate the performance of MB-pol with PIP, BPNN, 
and GAP representations of the short-range 2B and 3B energies and 
the corresponding effect on the  total interaction energies of the clusters.

\begin{figure}[!ht]
  \includegraphics[width=\textwidth]{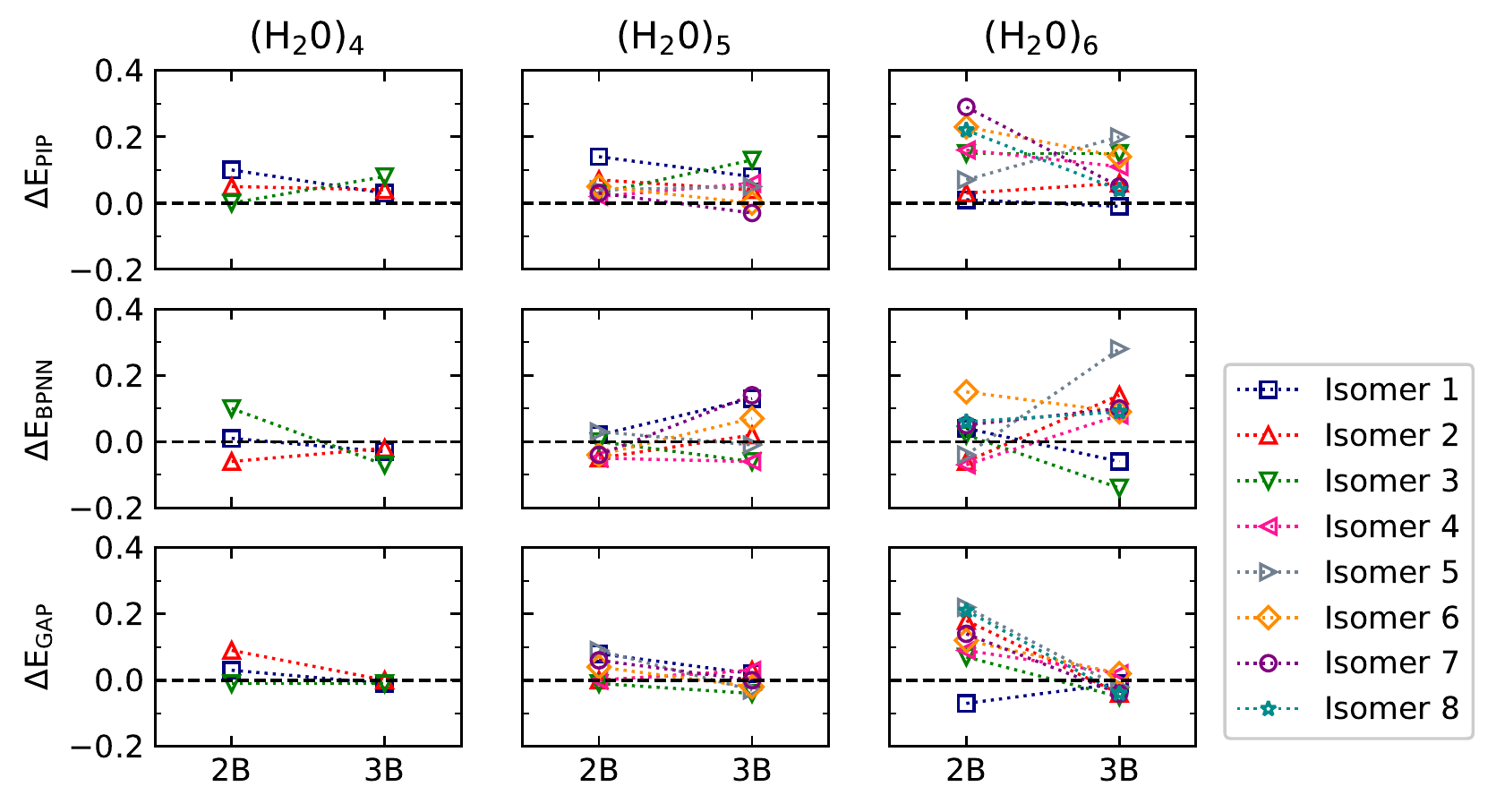}
  \caption{Errors (in kcal/mol) in the 2B and 3B interaction energies calculated 
  with PIP, BPNN and GAP short-range potentials with respect 
  to reference CCSD(T) values for water clusters (H$_2$O)$_n$, $n=4, 5, 6$.}
  \label{figure-5}
\end{figure}

An analysis of the 2B and 3B contributions to the total interaction energy 
of the water clusters is shown in Fig.\ \ref{figure-5}.
MB-pol errors with respect to the CCSD(T) reference values are smaller 
than 0.3\,kcal/mol in all cases, independent of the cluster size 
and geometry and independent of the approach that is used 
to represent the short-range 2B and 3B energies.
The errors increase somewhat with cluster size as the individual errors 
for the larger number of 2B and 3B terms can start to add up 
for cluster configurations that contain repeating dimer and trimer units.
This is mostly pronounced for 2B interaction energies.
While similar errors in 2B interaction energies are seen with the three potentials, GAP-MB-pol exhibits smaller errors in 3B interaction energies than PIP-MB-pol and BPNN-MB-pol.   

\begin{figure}[!ht]
  \includegraphics[width=6in]{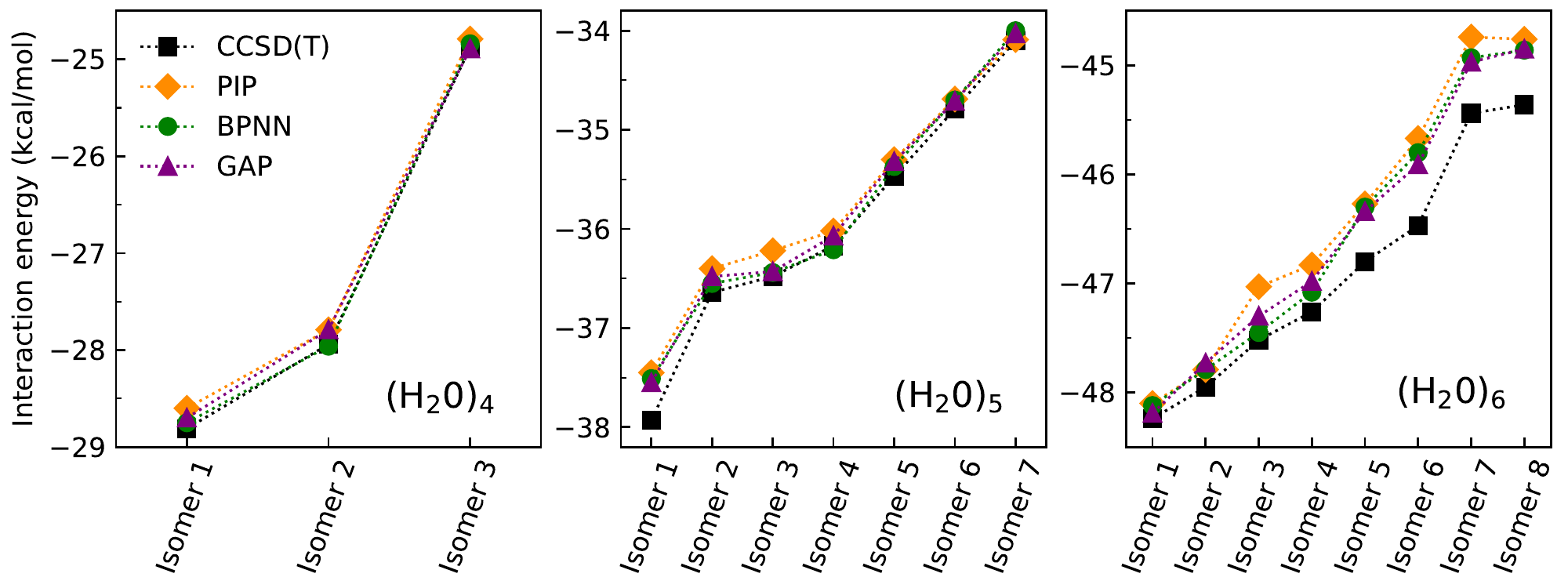}
  \caption{Interaction energies of the low-lying isomers of water clusters 
  (H$_2$O)$_n$, $n=4, 5, 6$, obtained using MB-pol with PIP, BPNN and GAP 
  short-range 2B and 3B potentials in comparison to CCSD(T)/CCSD(T)-F12b reference values.}
  \label{figure-6}
\end{figure}

Fig.\ \ref{figure-6} compares 
the total interaction energies of all water cluster isomers as obtained 
with MB-pol using PIP, BPNN, and GAP representations 
of short-range 2B and 3B energies in comparison 
to the CCSD(T)/CCSD(T)-F12b reference values.
In correspondence with the 2B and 3B contributions, 
the error in the total interaction energy increases with cluster size. 
Due to extended hydrogen bonding and symmetry, the ring-type isomers 
also have relatively large higher-body contributions that can be non-negligible 
and that can exhibit errors of similar magnitude as the 2B and 3B terms 
as has been shown in previous work\cite{reddy_sk_2016,medders_gr_2015a}.
The error for this type of isomers is thus particularly large. 
However, the  deviation in the computed interaction energies never exceeds 0.8\,kcal/mol 
and, most importantly, the relative order of the total interaction energies 
for the different isomers of each cluster is retained in all cases.
Overall we conclude that any of the investigated approaches 
to represent short-range 2B and 3B interaction energies within the MB-pol model 
is suitable to predict accurate interaction energies of water clusters.

\section{Conclusions}\label{sect:conclusions}
We have explored different representations of MB-pol short-range two-body (2B) and three-body (3B)
interaction energies using permutationally invariant polynomials (PIP), 
Behler-Parrinello neural networks (BPNN), and Gaussian approximation potentials (GAP). 
The accuracy of the three models has been assessed by comparing their ability to
reproduce large datasets of CCSD(T)/CBS 2B and 3B interaction energies
as well as in predicting the energetics of small water clusters,
which are always found to be within chemical accuracy (1 kcal/mol).
These results demonstrate that the three models are effectively equivalent, 
consistently exhibiting similar performance in representing many-body
interactions in water within the MB-pol framework. 
The most promising approach to further increase the accuracy for both the 2B and 3B terms involves increasing the number of reference calculations and optimizing the training set to cover more uniformly the relevant configuration space.
Our analysis of the 2B and 3B contributions to the MB-pol interaction energies can be taken as a case study for the general problem of the systematic construction of potentials derived from the many-body expansion. The combination between an accurate machine-learning representation of the short-range terms in combination with a physically sound form of long-range contributions provides a promising route to the development of accurate, efficient and transferable potential energy surfaces.

\section{Acknowledgements}
This work was supported by the National Science Foundation through grant no. ACI-1642336 (to F.P. and A.W.G.). 
This work used the Extreme Science and Engineering Discovery Environment (XSEDE), 
which is supported by National Science Foundation grant no. ACI-1548562. 
J.B. is grateful for a Heisenberg professorship funded by the DFG (Be3264/11-2).
E.Sz. would like to acknowledge the support of the Peterhouse Research Studentship and the support of BP International Centre for Advanced Materials (ICAM).
M.C. was supported by the European Research Council under the European Union's Horizon 2020 research and innovation programme (grant agreement no. 677013-HBMAP). G.I. acknowledges funding from the Fondazione Zegna.

\bibliography{biblio}

\end{document}